\documentclass{article}
\usepackage{graphicx}

\begin{document}

\title{Accelerating spaceships paradox and physical meaning of length contraction}
\author{Vesselin Petkov \\
Science College, Concordia University\\
1455 De Maisonneuve Boulevard West\\
Montreal, Quebec, Canada H3G 1M8\\
E-mail: vpetkov@alcor.concordia.ca}
\date{}
\maketitle

\begin{abstract}
A fifty-year old apparently paradoxical thought experiment involving two accelerating spaceships and a thread
that connects them states that the thread will break due to length contraction. This paradox still appears to be
regarded by some physicists as a proof that (i) only physical bodies, but not space, undergo relativistic length
contraction and (ii) a relativistically contracted body experiences a stress. This note, dedicated to the one
hundredth anniversary of Hermann Minkowski's paper ``Space and Time'', shows that a proper relativistic
treatment of the paradox demonstrates that it is a manifestation of a specific relativistic acceleration
phenomenon, which does not have a classical analog. This means that length contraction plays no role in the
resolution of the paradox. It is also shown that no stress is involved in the length contraction effect.
\end{abstract}

%01.55.+b, 03.30.+p
\vspace{.5cm}

In 1959 in a ``Note on Stress Effects due to Relativistic Contraction'' Dewan and Beran \cite{dewan59} (see
also \cite{evett60}-\cite{evett72}) proposed a thought experiment intended to demonstrate the existence of a
stress in a relativistically contracted body. In the seventies J. Bell revived that apparently paradoxical thought
experiment in a debate with colleagues at CERN. In 1976 he summarized his view in the paper ``How to teach
special relativity'' (reprinted in 1987 in \cite{bell}) and here is his formulation of the thought experiment \cite[p.
67]{bell}:

\begin{quotation}
Three small spaceships, $A$, $B$, and $C$, drift freely in a region of space remote from other matter, without
rotation and without relative motion, with $B$ and $C$ equidistant from $A$ (Fig. 1).

On reception of a signal from $A$ the motors of $B$ and $C$ are ignited and they accelerate gently... Let ships
$B$ and $C$ be identical, and have identical acceleration programmes. Then (as reckoned by an observer in
$A$) they will have at every moment the same velocity, and so remain displaced one from the other by a fixed
distance. Suppose that a fragile thread is tied initially between projections from $B$ and $C$. If it is just long
enough to span the required distance initially, then as the rockets speed up, it will become too short, because of
its need to Fitzgerald contract, and must finally break. It must break when, at a sufficiently high velocity, the
artificial prevention of the natural contraction imposes intolerable stress.
\end{quotation}

\begin{figure}[h]
\centering
\includegraphics[height=3cm]{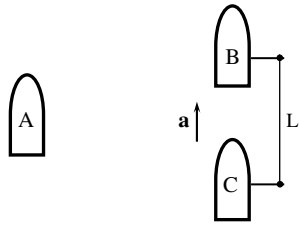}
\caption{Three spaceships $A$, $B$, and $C$ are initially at rest. At a given
moment in $A$'s reference frame $B$ and $C$, which are connected with a
thread, start to accelerate with the same proper acceleration. According to Bell,
the thread will break due to stress caused by length contraction. His conclusion
is based on the assumption that physical bodies contract relativistically, but
space does not.}
\label{bell1}
\end{figure}

Bell's motivation for reviving this thought experiment was to demonstrate the need to teach special relativity by
emphasizing ``the continuity with earlier ideas'' \cite[p. 67]{bell}. According to him the conclusion that the
thread would break ``is perfectly trivial in terms of A's account of things, including the Fitzgerald contraction''
\cite[p. 68]{bell}. His reasoning, which appears to be still shared by some physicists, is the following: the atoms
of the thread are deformed due to the distortion of their charges' fields and as a result the moving thread will
be shorter; as there are no charges and fields in empty space there is nothing that can cause the contraction of
space and therefore the distance between $B$ and $C$ will not contract (which appears consistent with the
condition that that distance should remain constant for $A$), but the thread will, which means that it will break.

The very formulation of the thought experiment itself -- that the distance between the accelerating spaceships
$B$ and $C$, as measured by $A$, remains constant -- appears to be paradoxical because according to special
relativity \textit{any} length should contract relativistically. As the distance between $B$ and $C$ must remain
constant for an observer in $A$ due to the same proper accelerations of $B$ and $C$ (which ensures that at
any moment in $A$, $B$ and $C$ have equal velocities relative to $A$), the only correct relativistic explanation
of the constant distance between $B$ and $C$ (from $A$'s point of view) is that the proper distance between
$B$ and $C$, measured by any of them, constantly increases as they accelerate \cite{boulware,lyle}. And the
distance between $B$ and $C$ must increase for $B$ and $C$ \textit{exactly as much as it decreases for} $A$
due to length contraction. As we will see this is precisely the case, which means that the constant increase of
the proper distance between $B$ and $C$, which are accelerating with the same proper accelerations, causes
the break of the thread, not length contraction. Such an increase of the distance between equally accelerating
bodies does not have a classical analog.

If it were the relativistic contraction of the thread, as explained by Bell, that caused its break, there would be a
real paradox. From $A$'s point of view the thread would break not because of the acceleration of $B$ and $C$,
as Bell argued, but due to length contraction. However, the proper distance between $B$ and $C$ should be
constant as required by the length contraction effect and therefore the break of the thread for $B$ and $C$
could not be explained. This can be demonstrated by taking into account (i) the \textit{relativity of motion} in
the length contraction effect and (ii) the \textit{reciprocity} of length contraction.

\begin{figure}[h]
\centering
\includegraphics[height=3cm]{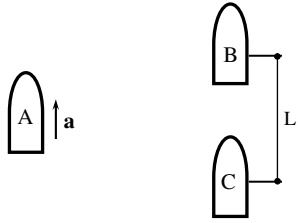}
\caption{Three spaceships $A$, $B$, and $C$ are initially at rest. But
unlike the thought experiment depicted in Fig.~\ref{bell1} it is spaceship $A$ that
starts to accelerate in this case. If the thread in Fig.~\ref{bell1} broke not because
of the acceleration of $B$ and $C$, but because of stress caused by length
contraction, then in the case depicted here an observer in $A$ would conclude
that the thread should also break. However, that would be a total mystery for
observers in $B$ and $C$ since $B$ and $C$ are stationary and the proper distance
between them remains the same, which also implies that the thread is subjected to no
stress. So taking into account the \textit{relativity} of motion in the length contraction
effect demonstrates that no stress is involved in this effect and therefore the
thread in Fig.~\ref{bell1} does not break because of length contraction.}
\label{bell2}
\end{figure}

Suppose that it is $A$ that accelerates, whereas $B$ and $C$ remain stationary (Fig.~\ref{bell2}). This means
that the proper distance between $B$ and $C$ (as measured by them) is constant since $B$ and $C$ do not
change their state of motion. As the speed of $A$ increases the length of the thread decreases relativistically as
monitored by an observer in $A$. If the \textit{distance} between $B$ and $C$ did not suffer relativistic length
contraction (as Bell supposed) an observer in $A$ would conclude that the thread would break. This means that
$B$ and $C$ should also observe the same event. But for them the break of the thread would be a miracle. The
correct explanation of the situation depicted in Fig.~\ref{bell2} is that from $A$'s point of view, both the
distance between $B$ and $C$ and the thread contract relativistically, which means that the thread is not
subjected to any stress and will not break. So, the thread will break in the original thought experiment shown in
Fig.~\ref{bell1}, but will remain perfectly intact in the situation represented in Fig.~\ref{bell2}. What makes the
two thought experiments different is the acceleration of the rockets -- $B$ and $C$ accelerate in
Fig.~\ref{bell1}, whereas it is $A$ that accelerates in Fig.~\ref{bell2}.

\begin{figure}[h]
\centering
\includegraphics[height=3cm]{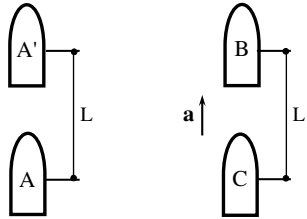}
\caption{Another spaceship $A^{\prime}$ is added to the thought experiment
depicted in Fig.~\ref{bell1}. Spaceships $A$ and $A^{\prime}$ are also connected
with a thread. If the thread in Fig.~\ref{bell1} broke not because of the acceleration of
$B$ and $C$, but because of stress caused by length contraction, then observers in $B$
and $C$ would conclude that the thread connecting $A$ and $A^{\prime}$ should also
break. However, that would be a total mystery for observers in $A$ and $A^{\prime}$
since the proper distance between $A$ and $A^{\prime}$ stays constant and the thread
suffers no stress. So taking into account the \textit{reciprocity} of length contraction
demonstrates that no stress is involved in this effect and therefore the thread in
Fig.~\ref{bell1} does not break because of length contraction.}
\label{bell3}
\end{figure}

That proper length remains constant in the length contraction effect can be also seen from its reciprocity as
shown in Fig.~\ref{bell3}. Consider a fourth spaceship $A^{\prime}$ which is at rest with respect to $A$. The
stationary spaceships $A$ and $A^{\prime}$ are also connected with a thread whose length remains constant
for observers in $A$ and $A^{\prime}$. As the speed of $B$ and $C$ increases the length of the thread
decreases relativistically as monitored by observers in $B$ and $C$. If the \textit{distance} between $A$ and
$A^{\prime}$ did not suffer relativistic length contraction (as Bell assumed) observers in $B$ and $C$ would
conclude that the thread would break. This means that observers in $A$ and $A^{\prime}$ should also observe
the same event. But for them the break of the thread would be a complete mystery since their state of motion
has not changed and the proper distance between them stayed the same. The correct explanation of the
situation depicted in Fig.~\ref{bell3} is that from $B$'s and $C$'s point of view, both the distance between $A$
and $A^{\prime}$ and the thread contract relativistically, which means that the thread is not subjected to any
stress and will not break. Again, what makes the situations represented in Fig.~\ref{bell1} and Fig.~\ref{bell3}
different is the acceleration of $B$ and $C$.

As for fifty years the accelerating spaceships paradox has been confusing some physicists and even relativists I
will briefly summarize the reasons which clearly demonstrate that (i) what causes the break of the thread
connecting $B$ and $C$ is their acceleration, not length contraction, and (ii) no stress is involved in the length
contraction effect. A proper relativistic treatment of this thought experiment shows that Bell's explanation is
based on a pre-relativistic intuition and as a result is incorrect on five counts.

1. Bell assumes that the thread contracts but space does not. However, both special relativity and the
experimental evidence, which confirmed length contraction, tell us unambiguously that everything -- physical
objects and space -- contract relativistically.  When we use the Lorentz transformations to determine the length
contraction between two points it does not matter whether these points are two points of space or are the end
points of the thread. The muon experiment \cite{muon,ellis}, which confirmed length contraction (in the rest
frame of the muon), demonstrated that space itself contracts relativistically.

2. Bell thought that the thread breaks as a result of a stress arising in it since a stress, according to him, is
always present when there is length contraction. But stress is an absolute (frame-independent) physical
quantity, which is represented by a tensor. This means that if the stress tensor is zero in one reference frame it
should be zero in all reference frames. Therefore, for observers in $B$ and $C$, who measure the constant
proper length of the thread in Fig.~\ref{bell2}, the stress tensor should be zero, which means that from $A$'s
viewpoint there should be no stress in the relativistically contracted thread either. This point, taken even alone,
demonstrates that no stress is involved in length contraction, which clearly indicates that length contraction does
not cause the break of the thread in Fig.~\ref{bell1}.

3. Bell insisted that the thread breaks due to length contraction -- the distance between $B$ and $C$ remains
constant for $A$ (space does not contract relativistically, according to Bell), but the thread contracts and
ultimately breaks. He apparently took it as self-evident that the thread did not break as a result of the
acceleration of $B$ and $C$. But as we saw above taking into account relativity of motion and the reciprocity of
length contraction demonstrates that the thread would not break, if only length contraction were involved in the
thought experiment.

4. If the reason for the break of the thread were length contraction, the proper length of the thread must be
constant as required by this effect and as we saw above, which is not the case in the discussed thought
experiment as Bell himself acknowledges: ``$B$, for example, sees $C$ drifting further and further behind, so
that a given piece of thread can no longer span the distance'' \cite[p. 68]{bell}. If the proper distance
increases, it is the acceleration of $B$ and $C$ that causes the break of the thread, not length contraction.

5. The very fact that this old paradox still manages to cause confusion demonstrates that Minkowski's view of
macroscopic physical bodies as four-dimensional worldtubes has not always been taken seriously. And indeed in
this paradox the authors who argued that there was a stress involved in a relativistically contracted body,
including Bell, assumed what appears to be self-evident -- that observers in $A$ measure the \textit{same}
three-dimensional thread that is measured by observers in $B$ and $C$, which means that the \textit{same}
three-dimensional thread is subjected to stress and deformation when it contracts. However, Minkowski pointed
out a hundred years ago that relativistic length contraction could not be understood as some kind of
deformation effect \cite[p. 81]{minkowski}:

\begin{quote}
According to Lorentz any moving body must have undergone a contraction in the direction in the direction of its
motion $[\ldots]$ This hypothesis sounds extremely fantastical, for the contraction is not to be looked upon as a
consequence of resistances in the ether, or anything of that kind, but simply as a gift from above, -- as an
accompanying circumstance of the circumstance of motion.
\end{quote}

Then Minkowski \cite[pp. 81-82]{minkowski} explained that the profound physical meaning of length contraction
is that while measuring the \textit{same} Lorentzian electron two observers in relative motion measure two
different three-dimensional cross-sections of the worldtube of the electron, which means that the observers
measure two \textit{different} three-dimensional electrons (what is `the same electron' is its worldtube).
Minkowski stressed that it is the fact that the observers measure \textit{two} three-dimensional cross-sections,
which have different lengths, that ``is the meaning of Lorentz's hypothesis of the contraction of electrons in
motion'' \cite[p. 82]{minkowski}.

In fact, it is Minkowski's attempt to reveal the physical meaning of relativity of simultaneity that led him to the
discovery of spacetime (length contraction is a specific manifestation of relativity of simultaneity) -- he realized
that relativity of simultaneity implies many spaces since a space is defined as one class of \textit{simultaneous}
events; hence relative simultaneity and therefore many spaces cannot exist in a three-dimensional world.
Minkowski wrote: ``We should then have in the world no longer space, but an infinite number of spaces,
analogously as there are in three-dimensional space an infinite number of planes. Three-dimensional geometry
becomes a chapter in four-dimensional physics'' \cite[pp. 79-80]{minkowski}.

Most probably, he took into account that a spatially extended body is defined in terms of \textit{simultaneity} --
as all parts of the body taken at a \textit{given} moment of time. Then, it does follow from relativity of
simultaneity that when measuring the \textit{same body} two observers in relative motion (who have different
classes of simultaneous events), in fact, measure \textit{two different three-dimensional bodies}. But this is
only possible if the worldtube of the body is a real four-dimensional object. In other words, the spaces of the
two observers intersect the worldtube at different angles and the three-dimensional cross-section of the
observer moving with respect to the body is shorter.

However, the two cross-sections have no \textit{objective} meaning since they are merely a \textit{description}
of the body's worldtube in terms of our three-dimensional language. This fact \textit{explains} why there is no
stress present in the relativistically contracted body -- the measured contracted body is not a real
three-dimensional object that is shortened and therefore deformed since there are no three-dimensional objects
in spacetime; what is measured by an observer moving with respect to the body (or more precisely, whose
worldtube forms an angle with the body's worldtube) is simply a three-dimensional cross-section whose length is
shorter than the length of the cross-section measured by an observer at rest with respect to the body. The two
cross-sections have \textit{equal} status -- they are \textit{imaginary} intersections of the observers' spaces
and the body's worldtube. Due to their equal status and since they do not represent real three-dimensional
objects in spacetime neither of the cross-sections involves any stress (an imaginary intersection cannot produce
real deformation and stress).

Minkowski's explanation of the physical meaning of length contraction is not just a \textit{different description}
as some relativists often say. It is \textit{the only description that does not contradict relativity}. Let me stress
-- no length contraction of a meter stick would be possible if the meter stick's worldtube did not exist as a
four-dimensional object. If the meter stick were a three-dimensional body, both observers would measure the
\textit{same} three-dimensional meter stick (the same set of simultaneously existing parts of the meter stick),
which would mean that the observers would have a common (absolute) class of simultaneous events in
contradiction with relativity \cite{petkov,petkov07}.

\begin{figure}[h]
\centering
\includegraphics[height=5.5cm]{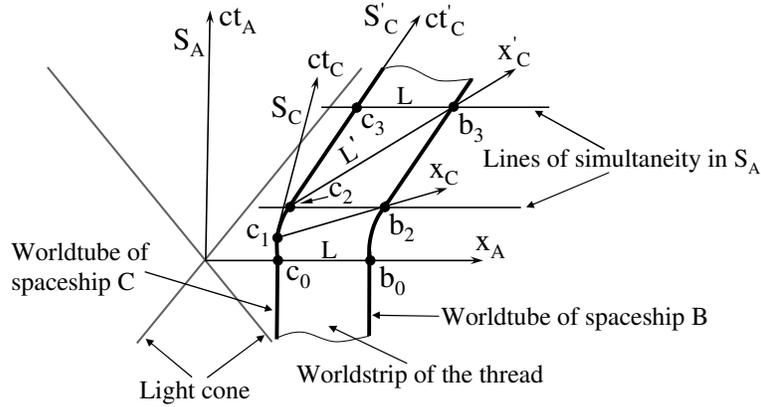}
\caption{The spaceships $B$ and $C$, represented by their worldtubes, start to
accelerate simultaneously in the reference frame $S_{A}$ of spaceship $A$ at
events $b_{0}$ and $c_{0}$, respectively. The wire connecting $B$ and $C$
is represented by its worldstrip. At event $c_{2}$ the wire breaks and the
motors of $B$ and $C$ stop simultaneously in $S_{A}$ at events $b_{2}$
and $c_{2}$, respectively; after that the spaceships continue to move
at a constant velocity relative to $S_{A}$. In $S_{A}$ the distance between
$B$ and $C$ (i.e. the distance between the simultaneous events $b_{2}$ and
$c_{2}$) is $L=x_{b_{2}}-x_{c_{2}}$. The instantaneously comoving inertial
reference frames $S_{C}$ and $S^{\prime}_{C}$ corresponding to the
events $c_{1}$ and $c_{2}$, respectively, are used to determine the increasing
proper distance $L^{\prime}$ between $B$ and $C$.} \label{bell4}
\end{figure}

Let me now show that it is the constantly increasing proper distance between the accelerating $B$ and $C$ that
(i) causes the break of the thread connecting them, and (ii) keeps the distance between $B$ and $C$ constant
as reckoned by an observer in $A$. To calculate the proper distance between $B$ and $C$ at different events
as measured by an observer in spaceship $C$ suppose that $B$ and $C$ are connected with a wire (instead of
a thread) whose diameter at its end, which connects it with $C$, is significantly smaller than the diameter of the
rest of the wire. So, when the wire breaks, it breaks at the end that connects it to $C$.

As shown in Fig.~\ref{bell4} the spaceships are represented by their worldtubes and the wire is represented by
its worldstrip. Along the path of $C$ there are devices in the inertial reference frame $S_{A}$ of spaceship $A$,
which monitor the break of the wire. Assume the wire breaks at event $c_{2}$. Then the device situated at the
closest distance to $c_{2}$ sends signals to $B$ and $C$ which shut down their motors simultaneously in
$S_{A}$. Despite that the signal sent to $B$ requires more time to reach $B$, it is possible to stop $B$'s and
$C$'s motors simultaneously in $S_{A}$; for instance, $C$'s motor can be stopped a little later after receiving
the shut down signal. In such a way $B$ and $C$ start moving simultaneously in $S_{A}$ at $t=t_{2}$ with
constant velocity with respect to $S_{A}$ (this is represented in the spacetime diagram by the straight
worldlines of $B$ and $C$ after the events $b_{2}$ and $c_{2}$, respectively).

In $S_{A}$ at moment $t=t_{0}$ the distance between spaceship $B$ and spaceship $C$ is the distance
between the simultaneous events $c_{0}$ and $b_{0}$:  $L=x_{b_{0}}-x_{c_{0}}$. As that distance remains
constant in $S_{A}$ the distance between the simultaneous events $c_{2}$ and $b_{2}$ at $t=t_{2}$, when
the wire breaks, is the same $L=x_{b_{2}}-x_{c_{2}}$. In the instantaneous comoving inertial frame at event
$c_{2}$ the proper distance between $B$ and $C$ is $L^{\prime}=x^{\prime}_{b_{3}}-x^{\prime}_{c_{2}}$.
The proper distance $L^{\prime}$ measured by observers in $B$ and $C$ is greater than the initial separation
$L$ between $B$ and $C$ and is also greater than the proper length
$L^{\prime}=x^{\prime}_{b_{3}}-x^{\prime}_{c_{2}}$ measured in the instantaneously comoving inertial
frame at event $c_{1}$. The constant increase of $L^{\prime}$ is a relativistic acceleration effect (with no
classical analog) resulting from the fact that the worldtubes of the accelerating spaceships $B$ and $C$ are
hyperbolae between the events $b_{0}$ and $c_{0}$ and $b_{2}$ and $c_{2}$, respectively. So, it is clearly
seen in Fig.~\ref{bell4} that the wire breaks because the proper distance between $B$ and $C$ increases, not
because of length contraction.

The assumption that $B$ and $C$ start moving at a constant velocity relative to $S_{A}$ at events $b_{2}$
and $c_{2}$ makes it possible to use Lorentz transformations to calculate the proper distance between $B$ and
$C$ as measured by an observer in $C$ at event $c_{2}$. The increase of the proper distance between $B$
and $C$ (as measured by any of them) in the case of constantly accelerating $B$ and $C$ is calculated in
\cite{lyle}.

In $S^{\prime}_{C}$ at the moment $t^{\prime}_{c_{2}}=t^{\prime}_{b_{3}}$ when the wire breaks the
\textit{proper} distance between $B$ and $C$ is $L^{\prime}=x^{\prime}_{b_{3}}-x^{\prime}_{c_{2}}$.
$L^{\prime}$ can be calculated by using the Lorentz transformations

\[
x^{\prime}_{b_{3}}=\gamma (x_{b_{3}}-vt_{b_{3}})
\]
\[
x^{\prime}_{c_{2}}=\gamma (x_{c_{2}}-vt_{c_{2}})
\]
where $\gamma=(1-v^{2}/c^{2})^{-1/2}$. As
$L^{\prime}=x^{\prime}_{b_{3}}-x^{\prime}_{c_{2}}$

\begin{equation}
L^{\prime}=\gamma (x_{b_{3}}-x_{c_{2}})-v\gamma (t_{b_{3}}-t_{c_{2}})
\label{Lp}
\end{equation}
In $S_{A}$ the $x$-coordinate of event $b_{3}$ is
\[
x_{b_{3}}=x_{c_{3}}+L
\]
where $L=x_{b_{3}}-x_{c_{3}}$ in $S_{A}$. But as
\[
x_{c_{3}}=x_{c_{2}}+v(t_{b_{3}}-t_{b_{2}})
\]
we can write
\[
x_{b_{3}}=x_{c_{2}}+L+v(t_{b_{3}}-t_{b_{2}}).
\]
Then
\begin{equation}
x_{b_{3}}-x_{c_{2}}=L+v(t_{b_{3}}-t_{c_{2}})
\label{xb3-xc2}
\end{equation}
where we took into account that $t_{b_{2}}=t_{c_{2}}$ in $S_{A}$. Now we can substitute (\ref{xb3-xc2}) in
(\ref{Lp}) and obtain the relation between $L^{\prime}$ and $L$ measured at $c_{2}$ in $S^{\prime}_{C}$
and $S_{A}$, respectively:
\[
L^{\prime}=\gamma L.
\]
Then the contracted length measured at $c_{2}$ in $S_{A}$ is

\begin{equation}
L_{contr}=\gamma^{-1}L^{\prime}=L.
\label{Lc}
\end{equation}

As the conceptual analysis demonstrated the proper distance between $B$ and $C$, measured by observers
there, increases exactly as much as the distance between $B$ and $C$, measured by an observer in $A$,
shortens due to length contraction. So, indeed the correct relativistic explanation of the constant distance
between $B$ and $C$, as reckoned by an observer in $A$, is that the proper distance between $B$ and $C$,
measured by any of them, increases. But the question of the \textit{physical meaning} of the increased proper
distance as measured by $B$ or $C$ remains unclear if it is assumed that the spaceships exist as
three-dimensional bodies -- as $B$ and $C$ accelerate with the \textit{same} proper acceleration the question
is: Why does the proper distance between them increase?

Taking into account the reality of the worldtubes of $B$ and $C$ and of the worldstrip of the wire provides a
complete explanation of the constant increase of the proper distance between $B$ and $C$ as measured by
any of them -- the spaces of the instantaneously comoving inertial reference frames corresponding to
consecutive events of the worldtube of $C$ intersect the existing worldtubes of $B$ and $C$ at different angles
as shown in Fig.~\ref{bell4}. At first sight it appears that the fact of the constant increase of the proper
distance between $B$ and $C$ as the spaceships continue to accelerate, explains why the wire breaks. But this
is not quite so. Due to the reality of the worldtubes of $B$ and $C$ and the worldstrip of the wire it follows that
the increasing proper distance, which is defined as the \textit{intersection} of the space of the instantaneously
comoving inertial frame at consecutive events, belonging to $C$'s worldtube, and the worldtubes of $B$ and
$C$, does not have any \textit{objective} meaning. This is best demonstrated by the fact that we can
\textit{choose} the \textit{same} non-orthogonal coordinate systems in \textit{all} inertial reference frames
($S_{A}$, $S_{C}$, and $S^{\prime}_{C}$), which means that the intersection defining the proper distance
between $B$ and $C$ depends on the convention used.

If the proper distance between $B$ and $C$ depends on our choice, how could the increase of something that
does not reflect anything objective cause the physical break of the wire? It seems to me the only remaining
explanation is the following. Before the beginning of the acceleration of $B$ and $C$ their worldtubes are
parallel as seen in Fig.~\ref{bell4} and the worldstrip of the wire attached to them is not stretched and does
not experience any stress. But starting at events $b_{0}$ and $c_{0}$ the two worldtubes are deformed which
means that the worldstrip of the wire is stretched by the deformation of the worldtubes of $B$ and $C$ and
ultimately broken. It should be stressed that nothing happens in spacetime -- simply at some time-like distance
from the events $b_{0}$ and $c_{0}$ the worldstrip of the wire is broken.

Had Bell taken seriously the reality of Minkowski spacetime and of worldtubes of macroscopic bodies he would
have realized that what causes the break of the wire is the permanently increasing proper distance between
$B$ and $C$ resulting from their acceleration, not because of length contraction. As it is the acceleration of $B$
and $C$ that breaks the wire, it would break even if there were no length contraction.

%\linebreak

\end{document}